\documentclass[conference]{IEEEtran}
\usepackage[utf8]{inputenc} 
\usepackage[T1]{fontenc}
\usepackage{url}
\usepackage{ifthen}
\usepackage{cite}
\usepackage[cmex10]{amsmath}
\usepackage{amssymb}% Use the [cmex10] option to ensure complicance
\usepackage{breqn}

\usepackage{hyperref}
\usepackage{algorithm}
\usepackage{algorithmic}
\usepackage{amsfonts}
\usepackage{epstopdf}
\usepackage{multirow}
\usepackage{subfigure}
\usepackage{changes}
\usepackage{flushend}
\definechangesauthor[name={me}, color=red]{Javad}
\definechangesauthor[name={you}, color=magneta]{you}
\usepackage{color}
\usepackage{tikz}
\usepackage[english]{babel}
\newtheorem{theorem}{Theorem}%[section]   % Theorems numbered by section
\newtheorem{definition}{Definition}%[section] % Definitions numbered by section
\newtheorem{lemma}{Lemma}%[section]       % Lemmas numbered by section
\newtheorem{proposition}{Proposition}%[section]
\newenvironment{proof}
  {\par\noindent\textit{Proof.}\quad}
  {\par\noindent\hfill$\square$}
\usepackage{graphicx}
\usepackage{nicematrix}
\newtheorem{remark}{Remark}
\usetikzlibrary{matrix}

\newcommand{\norm}[1]{\lVert#1\rVert}

\usepackage{mathtools}
\begin{document}
%\title{Explicit Hamiltonian Decompositions of Complete $k$-Uniform Hypergraphs and Its Applications \\ in Distributed Computing} 
\title{Constructing Hamiltonian Decompositions \\of Complete $k$-Uniform Hypergraphs} 

% %%% Single author, or several authors with same affiliation:
% \author{%
%  \IEEEauthorblockN{Author 1 and Author 2}
% \IEEEauthorblockA{Department of Statistics and Data Science\\
%                    University 1\\
 %                   City 1\\
  %                  Email: author1@university1.edu}% }

%%% Several authors with up to three affiliations:
\author{%
  \IEEEauthorblockN{Javad Maheri}
  \IEEEauthorblockA{Communication Systems Department \\
                    Eurecom\\
                    06410 Sophia Antipolis, France\\
                    Email: javad.maheri@eurecom.fr}
  \and
  \IEEEauthorblockN{Petros Elia}
  \IEEEauthorblockA{ Communication Systems Department\\                      Eurecom\\ 
                    06410 Sophia Antipolis, France\\ 
                    Email: elia@eurecom.fr}
           
    }
\maketitle

\begin{abstract}
% This paper presents a systematic framework for constructing explicit Hamiltonian decompositions of complete $k$-uniform hypergraphs $K_n^k$. Addressing limitations in prior work that relied primarily on non-constructive proofs, our approach provides explicit designs for a wide range of parameters $(n, k)$, covering cases where $n \mid \binom{n}{k}$ and \(n\)and \(k\) are prime numbers, and \(k <\frac{n}{2}\). Motivated by applications in distributed computing, our designs offer benefits in routing, load balancing, and fault tolerance. The construction is validated through rigorous proofs, illustrative examples, and algorithmic descriptions, demonstrating its practical and theoretical relevance.
Motivated by the wide-ranging applications of Hamiltonian decompositions in distributed computing, coded caching, routing, resource allocation, load balancing, and fault tolerance, our work presents a comprehensive design for Hamiltonian decompositions of complete \(k\)-uniform hypergraphs \(K_n^k\). Building upon the resolution of the long-standing conjecture of the existence of Hamiltonian decompositions of complete hypergraphs — a problem that was resolved using existence-based methods — our contribution goes beyond the previous explicit designs, which were confined to the specific cases of \(k=2\) and \(k=3\), by providing explicit designs for all \(k\) and \(n\) prime, allowing for a broad applicability of Hamiltonian decompositions in various settings.

\end{abstract}


\section{Introduction}
The Hamiltonian decomposition problem is a central subject in combinatorial optimization and graph theory. It involves partitioning the edge set of a graph into disjoint maximum-length (Hamiltonian) cycles, where each cycle traverses every vertex exactly once. 
 Consider for example the simplest case where we wish to partition the set of all $\binom{5}{2} = 10$ pairs of numbers from $1$ to $5$ (representing the edge set of the graph \(K_5\)), into $2$ subsets of $5$ pairs each, such that in each subset, each number (from $1$ to $5$) appears an equal number of times (twice) and each of the two subsets corresponds to a length-$5$ cycle. One such decomposition could be for instance 
 \(\{ \{1,2\}, \{2,3\}, \{3,4\}, \{4,5\}, \{5,1\}\}\) and \(\{\{1,3\}, \{2,4\}, \{3,5\}, \{4,1\}, \{5,2\}\}\), forming two maximum-length cycles.
 
This problem extends naturally to \(k\)-uniform hypergraphs, where each edge consists of \(k\) vertices (thus considering $k$-tuples instead of pairs). In this context, the concept of a Hamiltonian cycle extends in a natural way, as we recall later on. In this generalized setting, the problem becomes considerably more intricate, while maintaining substantial applicability in practical settings.
\subsection{Motivation and Applications}
%Coded Caching: Enhances data retrieval efficiency in distributed storage systems, uncoded Map reduce, GDD design
Hamiltonian decompositions find their way in various coded caching solutions (cf.,~\cite{wan2021combination},\cite{8262760}) as well as can relate to multiple other caching \cite{10206719,10402116},\cite{wan2020index}, \cite{yan2021fundamental}, \cite{10619106}, \cite{engelmann2017content}, \cite{lampiris2020full} and distributed computing problems~\cite{8437333}, \cite{10458969}, \cite{khalesi2023multi}, \cite{malak2024multi}, \cite{10694316}. Furthermore, in distributed networks, such decompositions often ensure a structured traversal of nodes, thus relating to more efficient routing, task scheduling and resource allocation, while often also guaranteeing reduced communication overheads~\cite{shi2009hamiltonian},\cite{bermond1998hamilton}. 
 Hamiltonian cycles have also been linked to efficient failure recovery and energy efficiency in sensor networks~\cite{kannan2004game,5934919}, as well as to scalability, robustness and efficiency in various distributed systems~\cite{wu2009system},\cite{wu2008optimizing}. Additional benefits of designs associated to Hamiltonian decompositions have also been discovered in the context of load balancing and fault tolerance~\cite{hu2020fault}.

Despite such broad applicability, explicit constructions of Hamiltonian decompositions of $k$-uniform hypergraphs remain very restricted, mainly because the considerable progress in proving the existence of such Hamiltonian decompositions, has almost always employed existence-type proofs. To date, explicit designs of Hamiltonian decompositions have been limited to the cases of $k=2$, $k=3$, and a specific instance of $k=4$ (corresponding to $n=9$).
% particularly for large parameters. Existing work has largely focused on existential proofs, leaving a gap in practical, constructible designs. 
\subsection{Prior Work}

The problem of Hamiltonian decompositions of hypergraphs has been studied extensively, with the pioneering work of \cite{bermond1978hamiltonian} conjecturing that the complete \(k\)-uniform hypergraph\footnote{A complete \( k \)-uniform hypergraph with \( n \) nodes is a hypergraph where every subset of \( k \) nodes forms a hyperedge, resulting in \(\binom{n}{k}\) hyperedges.} \(K_n^{k}\) has a Hamiltonian decomposition into Berge cycles\footnote{See a detailed description later on, as well as see \cite{bailey2010hamiltonian} and \cite{saengchampa2024hamiltonian}.} if \(n \mid {n \choose k}\). This conjecture and the realization of its practical significance, sparked a flurry of efforts for resolving it, with notable works found in \cite{bermond1978hypergraphes},\cite{verrall1994hamilton}%-\cite{petecki2014cyclic}
, all employing existence results. The conjecture was conclusively resolved in \cite{kuhn2014decompositions} which revealed that there exists a Hamiltonian \(k\)-uniform hypergraph decomposition into Hamiltonian Berge cycles, for all applicable cases (where the divisibility condition \( n\mid { n \choose k}\) applies). To solve this problem, the work in~\cite{kuhn2014decompositions} considered a complete bipartite graph \(G\) in the context of the work by Tillson et al. in~\cite{tillson1980hamiltonian} and showed that if \(G\) has a perfect matching, then a Hamiltonian decomposition can be found for \(K_n^k\). This approach was of an existence, rather than of a constructive nature, as it directly applied the theorem by Kruskal \cite{kruskal1963number} and Katona \cite{katona1987theorem} to ensure that Hall's condition \cite{hall1935} was satisfied for \(G\).

In terms of constructive solutions, Walecki in \cite{alspach2008wonderful} provided the first explicit construction for Hamiltonian decomposition in simple complete graphs \(K_n\) (\(k=2\)), doing so for odd \(n\). 
For the more involved case of hypergraphs (\(k>2\)), some instances of the problem for \(k=3\) were addressed in %\textcite{verrall1994hamilton} and \textcite{bermond1978hamiltonian}. Subsequently, 
\cite{bailey2010hamiltonian}, which first showed that a Hamiltonian decomposition of \(K_{n}^{k}\) is equivalent to a collection of 1-\((n, k, k)\) designs
\cite{khosrovshahi2006large}, and which then employed a clique-finding approach \cite{ostergaard2005constructing} alongside a difference-pattern approach \cite{bailey2010hamiltonian}, to constructively provide a decomposition for the cases of 
\(k = 3\) and \(n \leq 16\), as well as for the case of \(K_9^4\). Additionally, again for the case of \(k=3\), the work in~\cite{meszka2009decomposing} modified the aforementioned difference-pattern method to constructively resolve the cases of \(n<32\), while later the work in \cite{huo2015jirimutu} extended this to \(n<46\) (except for \(n = 43\)). To date, apart for \(k=2,3\), and the specific case of \(K_9^4\), no explicit constructions are known for Hamiltonian decompositions of complete \(k\)-uniform hypergraphs.% into Hamiltonian Berge cycles.

Our work here introduces a construction for Hamiltonian decompositions of \(K_n^k\), for all prime \(n\) and \(k\), \(k < \frac{n}{2}\), and does so using a generator-based framework. 

\subsection{Notations}
%\vspace{5pt}
For an integer \(n\), the set of integers from \(1\) to \(n\) is denoted by \([n] \triangleq \{1, 2, \dots, n\}\). For any integer \(k \leq n\), we use \({[n] \choose k}\) to represent the set of all \(k\)-sized subsets of \([n]\). Sets or \(k\)-tuples are represented using calligraphic letters with curly braces, such as \(\mathcal{A} = \{a_1, a_2, \dots, a_k\}\). Ordered sets and vectors are denoted with underlined small letters, for example, \(\underline{a} = [a_1, a_2, \dots, a_k]\). Hamiltonian cycles are represented by bold calligraphic letters, e.g., \(\boldsymbol{\mathcal{S}} = (v_1, \mathcal{E}_1, v_2, \dots, v_n, \mathcal{E}_n, v_1)\), where \(v_i\) are vertices and \(\mathcal{E}_i\) are hyperedges (\(k\)-sized sets that may or may not be ordered) leading from the previous to the next vertex. Sets of sets are denoted by bold uppercase letters, such as \(\mathbf{A}\), while sets of vectors are represented using underlined bold uppercase letters, such as \(\underline{\mathbf{A}}\). The cardinality of any set \(\mathcal{A}\)  is denoted by \(|\mathcal{A}|\). Finally, \(\norm{\underline{a}}_1 \triangleq |a_1| + |a_2| + \dots + |a_k|\) and \(\norm{\underline{a}}_\infty \triangleq \max_{1 \leq i \leq k} |a_i|\), respectively represent the 1-norm and infinity norm of a vector \(\underline{a} = [a_1, a_2, \dots, a_k]\).
    % starting point does not matter, circular
    % \item \textbf{Vectors:} Vectors are represented by underlined lowercase letters, e.g., $\underline{v} = [v_1, v_2, \dots, v_k]$.
\section{Preliminaries}
%\vspace{5pt}
\subsection{Definitions}
%\vspace{5pt}
A complete $k$-uniform hypergraph $K_n^k$ consists of a vertex set $\mathcal{V}=\{v_1, v_2, \ldots, v_n\}$ and an edge set $\mathbf{E}$ where each hyperedge \(\mathcal{E} \in \mathbf{E} \) corresponds to a $k$-sized subset of $\mathcal{V}$.
    %i.e. \(\mathbf{E}={[n] \choose k}\).
%\end{itemize}
The total number of hyperedges is $|\mathbf{E}| = \binom{n}{k}$.

Following Berge's definition \cite{berge1970graphes}, a Hamiltonian cycle 
\begin{small}
\begin{equation}
\boldsymbol{\mathcal{S}}=(v_1,\ \mathcal{E}_1,\ v_2,\ \mathcal{E}_2,\ \dots,\ \mathcal{E}_{n-1},\ v_n,\ \mathcal{E}_n,\ v_1)
\end{equation}
\end{small}
of the $k$-uniform hypergraph $K_n^k$, satisfies the following:
\begin{itemize}
    \item The hyperedges $\mathcal{E}_1, \dots, \mathcal{E}_n \in \mathbf{E} $ are distinct.
    %(\(k\)-sized sets or \(k\)-tuples).
    \item Each hyperedge $\mathcal{E}_i$ contains $v_{i+1}$ and $v_i$, where the indices are modulo \(n\).
     % \item The subsequence \((v_1, v_2, \dots, v_n)\) is a permutation of Vertex set \(V\).\\
\end{itemize}
We proceed with the definition of the Hamiltonian decomposition.
\begin{definition} \label{def:Berge}
A Hamiltonian decomposition of $K_n^k$ partitions the edge set $\mathbf{E}$ into $N = \frac{{n \choose k}}{n}$ disjoint subsets $\ \mathbf{U}_1,\ \ldots,\mathbf{U}_N$, where each subset \(\mathbf{U}_i\) has \(n\) hyperedges \(\mathcal{U}_{i,1}, \mathcal{U}_{i,2} \dots, \mathcal{U}_{i,n} \in \mathbf{U}_i\) and forms the Hamiltonian cycle
\begin{small}
    \begin{equation}
        \boldsymbol{\mathcal{S}_i}=(v_1,\ \mathcal{U}_{i,1},\ v_2,\ \mathcal{U}_{i,2},\ \dots,\ \mathcal{U}_{i,n-1},\ v_n,\ \mathcal{U}_{i,n},\ v_1). 
    \end{equation}
\end{small} 
\end{definition}
Naturally, it is required that \(n\) divides \(\binom{n}{k}\). 

The above decomposition problem can be reformulated as the problem of partitioning the set \(\mathbf{A}_{n,k} = {[n] \choose k}\) of all \(k\)-tuples from \([n]\) into \(N = \frac{{n \choose k}}{n}\) subsets \( \mathbf{U}_1,\mathbf{U}_2,\ \dots,\ \mathbf{U}_N\), such that each subset \(\mathbf{U}_i\) consists of \(n\) distinct \(k\)-tuples \(\mathcal{U}_{i,m} \in \mathbf{U}_i, \ m \in [n]\), forming a Hamiltonian cycle as in Definition~\ref{def:Berge}.

As we will see later on, it will become easier to construct Hamiltonian cycles from specifically ordered sets (vectors). Thus we will henceforth consider the vector set \(\underline{\mathbf{A}}_{n,k} \triangleq \{\pi_{\mathcal{A}}(\mathcal{A})  \ : \  \mathcal{A} \in \mathbf{A}_{n,k} \} \)  of vectors \(\underline{a} = \pi_{\mathcal{A}}(\mathcal{A})\), whose elements though are carefully ordered. The details of this ordering and of the corresponding permutations $\pi_{\mathcal{A}}$ will be discussed in Subsection \ref{subsec:VO}.  
For now, let us simply consider \(\underline{\mathbf{A}}_{n,k}\) to be the set of all $k$-length vectors with different numbers from $1$ to $n$, except that the entries of each vector are ordered in a manner that will facilitate our design later on.  Finally, let us note that we consider two vectors to be \emph{distinct} if no rotation (cyclic shift) of the first vector can render it identical to the second. 

In this context, a Hamiltonian cycle takes the following form. 

\begin{definition}
\label{def:cycle}
%    A Hamiltonian cycle from \(\underline{\mathbf{A}}_{n,k}\) takes the form
    A Hamiltonian cycle takes the form
    \begin{small}
        
    \begin{equation}
        \boldsymbol{\mathcal{S}} = (v_1, \underline{a}_1, v_2, \underline{a}_2, \dots, v_n, \underline{a}_n, v_1)
    \end{equation}
    \end{small}
    \!\! where the edge set \(\underline{\mathbf{C}} = \{\underline{a}_1, \underline{a}_2, \dots, \underline{a}_n\} \subset \underline{\mathbf{A}}_{n,k}\) is a set of $n$ distinct $k$-length vectors, such that:
    \begin{itemize}
        % \item the vectors \(\underline{a}_1, \underline{a}_2, \dots, \underline{a}_n \) are invariant under cyclic shifts.
        \item \(v_1, v_2, \ldots, v_n\) are distinct.
        \item each \( \underline{a}_i \) is of the form \([v_{i}, v_{i+1}, \dots]\) (indices modulo \(n\)).
        \item \((v_1, v_2, \dots, v_n)\) is a permutation of alphabet set \([n]\).
        % no permutation of set\{1,2, \dots, n\}
        % no vertices
        % the point is we give the cycle a form
        % orbit Method
        % drawing for a hamilton cycle
        % a_i s are ordered, 
    \end{itemize}
\end{definition}

\subsection{Adopted ordering}
\label{subsec:VO}
Given an arbitrary \(k\)-tuple \(\mathcal{A} =\{a_1,\ a_2,\ \dots ,\ a_{k-1},\ a_k\} \in \mathbf{A}_{n,k}\) such that \(a_1 <a_2<\dots<a_k\), we define the vector \(\underline{a}=\pi_{\mathcal{A}}(\mathcal{A})=\pi_t([a_1,a_2, \dots, a_k]) = [a_{1+t},\ a_{2+t},\ \dots,\ a_{t+k-1},\ a_{t}]\) (indices modulo \(k\)) as the adopted ordering of \(\mathcal{A}\), where \(\pi_t(.)\) rotates \([a_1, a_2, \dots, a_k]\) \(t\) times such that
the corresponding difference vector \(\underline{a}\) defined as
% for the representatives , r_g are valid permutation
\begin{small}
\begin{equation}
 \underline{d}_{\underline{a}} = [d_{ro}^n(a_{2+t}, a_{1+t}),\ \dots,\ d_{ro}^n(a_{t},a_{t+k-1})] 
 \end{equation}
 \end{small}
(with \(d_{ro}^n(y, x) = y - x\) if \(y \geq x\), and \(d_{ro}^n(y, x) = n - (x - y)\) otherwise) belongs to \(\{1, \dots , \left\lfloor\frac{n}{2}\right\rfloor\}^{k-1}\)
%We define also the vector \(\pi_{\mathcal{A}}(\mathcal{A})=\pi_t([a_1,a_2, \dots, a_k]) = [a_{1+t},\ a_{2+t},\ \dots,\ a_{k+t}]\) (indices modulo \(k\)) as the adopted ordering of \(\mathcal{A}\) , where \(\pi_t(.)\) rotates \([a_1, a_2, \dots, a_k]\) \(t\) times where 
and the value \(t\) must be chosen such that \(d_{ro}^n(a_{1+t}, a_{t})\geq\norm{\underline{d}_{\underline{a}}}_\infty.\) This
choice satisfies
\begin{small}
\begin{equation}
\label{eq:condition}
\norm{\underline{d}_{\underline{a}}}_{\infty} \leq n-\norm{\underline{d}_{\underline{a}}}_1
\end{equation}
\end{small}
where 
%Here, \(\norm{\underline{d}_{\underline{a}}}_1\) is 
\begin{small}
\begin{equation}
\label{eq:norm1}
\norm{\underline{d}_{\underline{a}}}_1 = \sum_{j=1}^{k-1} \left|d_{ro}^n(a_{t+j+1}, a_{t+j})\right|= \sum_{j=1}^{k-1}d_{ro}^n(a_{t+j+1}, a_{t+j})
\end{equation}
\end{small}
and 
\begin{small}
\begin{equation}
\norm{\underline{d}_{\underline{a}}}_{\infty} \!=\!\! \! \max_{1 \leq j \leq k-1} \left|d_{ro}^n(a_{j+1+t}, a_{j+t})\right|\!=\!\!\! \max_{1 \leq j \leq k-1} d_{ro}^n(a_{t+j+1}, a_{t+j}).
\end{equation}
\end{small}

\subsection{Examples}%: \(\underline{\mathbf{A}}_{7,3}\)}
%illustrating exmp
\label{exp:HD}
%\vspace{5pt}
For the case of \(n=7\) and \(k=3\), we first consider the complete hypergraph defined by the set $\underline{\mathbf{A}}_{7,3}$ of all ${7 \choose 3} = 35$ distinct ordered triplets. We consider the following decomposition consisting of 
%\[ A_{7,3} = \{(1,2,3), (1,2,4), (1,2,5), (1,2,6), (1,2,7), (1,3,4),\]\[ (1,3,5), (1,3,6), (1,3,7), (1,4,5), (1,4,6),(1,4,7), (1,5,6),\]\[ (1,5,7), (1,6,7), (2,3,4), (2,3,5), (2,3,6), (2,3,7), (2,4,5), \]\[ (2,4,6),(2,4,7), (2,5,6), (2,5,7), (2,6,7), (3,4,5), (3,4,6), \]\[ (3,4,7), (3,5,6), (3,5,7), (3,6,7), (4,5,6), (4,5,7), (4,6,7),\]\[  (5,6,7)\} \]
 \(N = \frac{35}{7} = 5\) sets \(\underline{\mathbf{U}}_1, \ \underline{\mathbf{U}}_2,\ \underline{\mathbf{U}}_3,\ \underline{\mathbf{U}}_4,\ \underline{\mathbf{U}}_5\) taking the form%\footnote{We will here remove the commas separating the elements of the vectors. Thus in our example, the element $[123]$ refers to the vector $[1,2,3]$. We will use this shorter notation in examples We will do so whenever no confusion arises from such removal. }:
\begin{small}
    \begin{align*}
\underline{\mathbf{U}}_1 \!= \!& \{ [1,2,3], [2,3,4], [3,4,5], [4,5,6], [5,6,7], [6,7,1], [7,1,2]\},    \\
\underline{\mathbf{U}}_2 \!= \!& \{[1,2,4], [2,3,5], [3,4,6], [4,5,7], [5,6,1], [6,7,2], [7,1,3]\}, \\
\underline{\mathbf{U}}_3 \!= \!& \{[1,3,4], [2,4,5], [3,5,6], [4,6,7], [5,7,1], [6,1,2], [7,2,3]\}, \\
\underline{\mathbf{U}}_4 \!= \!& \{[1,3,5], [2,4,6], [3,5,7], [4,6,1], [5,7,2], [6,1,3], [7,2,4]\}, \\
\underline{\mathbf{U}}_5 \!= \!& \{[1,2,5], [2,3,6], [3,4,7], [4,5,1], [5,6,2], [6,7,3], [7,1,4]\}.
\end{align*}
\end{small}
and we verify that \(\underline{\mathbf{U}}_i \cap \underline{\mathbf{U}}_j = \emptyset\) for \(i \neq j\) and \(\underline{\mathbf{U}}_1 \cup \underline{\mathbf{U}}_2 \cup \underline{\mathbf{U}}_3 \cup \underline{\mathbf{U}}_4 \cup \underline{\mathbf{U}}_5 = \underline{\mathbf{A}}_{7,3}\) such that each \(\underline{\mathbf{U}}_i\) contains \(n=7\) distinct ordered triplets. Finally we verify that each \(\underline{\mathbf{U}}_i\) corresponds to a Hamiltonian cycle \(\boldsymbol{\mathcal{S}}_i\) (cf.~Definition \ref{def:cycle}) of the form
\begin{small}
%     \begin{align*}
% \boldsymbol{\mathcal{S}}_1 \!= \!& ( 1, [1,2,3], 2, [2,3,4], 3, [3,4,5], 4, [4,5,6], 5, [5,6,7], 6, [6,7,1], 7, [7,1,2], 1]    \\
% \boldsymbol{\mathcal{S}}_2 \!= \!& (1, [1,2,4], 2, [2,3,5], 3, [3,4,6], 4, [4,5,7], 5, [5,6,1], 6, [6,7,2], 7, [7,1,3], 1] \\
% \boldsymbol{\mathcal{S}}_3 \!= \!& (1, [1,3,4], 3, [3,5,6], 5, [5,7,1], 7, [7,2,3], 2, [2,4,5], 4, [4,6,7], 6, [6,1,2], 1] \\
% \boldsymbol{\mathcal{S}}_4 \!= \!& (1, [1,3,5], 3, [3,5,7], 5, [5,7,2], 7, [7,2,4], 2, [2,4,6], 4, [4,6,1], 6, [6,1,3], 1 ] \\
% \boldsymbol{\mathcal{S}}_5 \!= \!& (1, [1,2,5], 2, [2,3,6], 3, [3,4,7], 4, [4,5,1], 5, [5,6,2], 6, [6,7,3], 7, [7,1,4], 1].
% \end{align*}
\[ \boldsymbol{\mathcal{S}}_1 =  ( 1, [1,2,3], 2, [2,3,4], 3, [3,4,5], 4, [4,5,6], 5, [5,6,7], 6, \]\[ [6,7,1], 7, [7,1,2], 1),\]  
\[\boldsymbol{\mathcal{S}}_2 = (1, [1,2,4], 2, [2,3,5], 3, [3,4,6], 4, [4,5,7], 5, [5,6,1], 6,\]\[ [6,7,2], 7, [7,1,3], 1),\] 
\[\boldsymbol{\mathcal{S}}_3 = (1, [1,3,4], 3, [3,5,6], 5, [5,7,1], 7, [7,2,3], 2, [2,4,5], 4, \]\[[4,6,7], 6, [6,1,2], 1),\]
\[\boldsymbol{\mathcal{S}}_4 =  (1, [1,3,5], 3, [3,5,7], 5, [5,7,2], 7, [7,2,4], 2, [2,4,6], 4, \]\[[4,6,1], 6, [6,1,3], 1 ),\ \] 
\[\boldsymbol{\mathcal{S}}_5 =  (1, [1,2,5], 2, [2,3,6], 3, [3,4,7], 4, [4,5,1], 5, [5,6,2], 6, \]\[ [6,7,3], 7, [7,1,4], 1).\]
\end{small}
%\subsection{Example: \(\underline{\mathbf{A}}_{8,3}\)}
%\label{exp2}
%\vspace{5pt}

For the case of \(n=8\) and \(k=3\), we consider the complete hypergraph defined by the set $\underline{\mathbf{A}}_{8,3}$ of all ${8 \choose 3} = 56$ properly ordered triplets. If we partition \(\underline{\mathbf{A}}_{8,3}\) into $\underline{\mathbf{U}}_1,\underline{\mathbf{U}}_2,\dots, \underline{\mathbf{U}}_7$ which respectively take the form:
 \begin{small}
\begin{align*}
\{ [1,2,3],\! [2,3,4],\! [3,4,5],\! [4,5,6],\! [5,6,7],\! [6,7,8],\! [7,8,1],\![8,1,2]\},\\
\{[1,3,4],\! [2,4,5],\! [3,5,6],\! [4,6,7],\! [5,7,8],\! [6,8,1],\! [7,1,2],\![8,2,3]\},\\
\{[1,2,4],\! [2,3,5],\! [3,4,6],\! [4,5,7],\! [5,6,8],\! [6,7,1],\! [7,8,2],\![8,1,3]\},\\
\{[1,3,5],\! [2,4,6],\! [3,5,7],\! [4,6,8],\! [5,7,1],\! [6,8,2],\! [7,1,3],\![8,2,4]\},\\
\{[1,4,5],\! [2,5,6],\! [3,6,7],\! [4,7,8],\! [5,8,1],\! [6,1,2],\! [7,2,3],\![8,3,4]\},\\
\{[1,2,5],\! [2,3,6],\! [3,4,7],\! [4,5,8],\! [5,6,1],\! [6,7,2],\! [7,8,3],\![8,1,4]\},\\
\{[1,3,6],\! [2,4,7],\! [3,5,8],\! [4,6,1],\! [5,7,2],\! [6,8,3],\! [7,1,4],\![8,2,5]\}
\end{align*} 
%\begin{dmath}
% \underline{\mathbf{U}}_1 = \{ [1,2,3], [2,3,4], [3,4,5], [4,5,6], [5,6,7], [6,7,8], [7,8,1],[8,1,2]\}, \underline{\mathbf{U}}_2 = \{[1,3,4], [2,4,5], [3,5,6], [4,6,7], [5,7,8], [6,8,1], [7,1,2],[8,2,3]\},\end{dmath}
 %\[
% \underline{\mathbf{U}}_1 = \{ [1,2,3], [2,3,4], [3,4,5], [4,5,6], [5,6,7], [6,7,8], [7,8,1]\]\[,[8,1,2]\},\]
%\[\underline{\mathbf{U}}_2 = \{[1,3,4], [2,4,5], [3,5,6], [4,6,7], [5,7,8], [6,8,1], [7,1,2]\]\[,[8,2,3]\},\]
% \[\underline{\mathbf{U}}_3 = \{[1,2,4], [2,3,5], [3,4,6], [4,5,7], [5,6,8], [6,7,1], [7,8,2]\]\[,[8,1,3]\},\]
% \[\underline{\mathbf{U}}_4 =\{[1,3,5], [2,4,6], [3,5,7], [4,6,8], [5,7,1], [6,8,2], [7,1,3]\]\[,[8,2,4]\},\]
% \[\underline{\mathbf{U}}_5 = \{[1,4,5], [2,5,6], [3,6,7], [4,7,8], [5,8,1], [6,1,2], [7,2,3]\]\[,[8,3,4]\},\]
% \[\underline{\mathbf{U}}_6 =\{[1,2,5], [2,3,6], [3,4,7], [4,5,8], [5,6,1], [6,7,2], [7,8,3]\]\[,[8,1,4]\},\]
% \[\underline{\mathbf{U}}_7 =\{[1,3,6], [2,4,7], [3,5,8], [4,6,1], [5,7,2], [6,8,3], [7,1,4]\]\[,[8,2,5]\}\] 
\end{small}
we note that $\underline{\mathbf{U}}_4$ entails two shorter cycles (of length $4$) 
\begin{small}
\[
(1,[1,3,5],3, [3,5,7], 5, [5,7,1], 7, [7,1,3], 1),\]  
\[
(2, [2,4,6],4, [4,6,8], 6,[6,8,2], 8,[8,2,4],2)\] 
\end{small}
\hspace{-6pt} which means that this partition \(\underline{\mathbf{U}}_1, \dots, \underline{\mathbf{U}}_7\) is not a Hamiltonian decomposition.

\section{Designing the Hamiltonian Decomposition}
%\vspace{10pt}
Before describing the design, we provide some preliminary definitions that are needed. 
\subsection{Generator-Based Framework}
We will use the notation $\underline{g}$ to refer to a \emph{generator} vector with $k$ integer entries $\underline{g}= [g_1, g_2, \ldots, g_k]$. For design purposes, we will be considering only generators $\underline{g}$ for which $g_i\geq 1, \forall i \in [k-1]$ and \(g_1+\dots+g_{k-1}\leq n-1\). For each $\underline{g}$ we consider the so-called \emph{representative} vector  \
\begin{small}
    \begin{equation}
          \underline{r}_{\underline{g}} = [1, 1 + g_1, 1 + g_1 + g_2, \ldots, 1 + g_1 + \cdots + g_{k-1}]
    \end{equation}
\end{small}
as well as the subsequent vector set 
\begin{small}
\begin{equation} \label{eq:Cg}
     \underline{\mathbf{C}}_{\underline{g}}= \{\underline{r}_{\underline{g}} + p\cdot \underline{1} \ (\text{mod } n) ,\  p= 0,1, \dots,n-1\}
\end{equation}
\end{small}
of all cyclic shifts of \(\underline{r}_{\underline{g}}\), together with its corresponding cycle \begin{small}
\begin{equation} \label{eq:cycleS1}
     \boldsymbol{\mathcal{S}}_{\underline{g}}= (1, \underline{r}_{\underline{g}}, 1+g_1,  \underline{r}_{\underline{g}}+g_1\cdot \underline{1}, 1+2g_1, \underline{r}_{\underline{g}}+2g_1\cdot \underline{1}, \dots ).
     % we donot know if it is a Hamiltonian or not. and also S_g must be belonged to C_g
\end{equation}
\end{small}
In the above, \(\underline{1} = [1, 1, \dots, 1]\) is the all $1$ vector of length $k$. Naturally the modulo operation ensures that all entries are integers from \([1,n]\).

We will be designing generators $\underline{g}$ that yield cycles \(\boldsymbol{\mathcal{S}}_{\underline{g}}\) that are Hamiltonian.  Our design is explicit, in the sense that we will be describing the entire set of generators, and thus the corresponding Hamiltonian cycles. We first need the following definitions.
%\subsection{Hamiltonian Cycles from Representatives}
%\vspace{5pt}
%A representative $\underline{r}_{\underline{g}}$ corresponding to the generator \(\underline{g} \in \underline{G}\) produces the set \(\underline{\mathbf{C}}_{\underline{g}}\) which is the collection of  all cyclic shifts of \(\underline{r}_{\underline{g}}\), where the shift is determined by the scalar \(p\),
%
%\begin{small}
%\begin{equation}
%     \underline{\mathbf{C}}_{\underline{g}}= \{\underline{r}_{\underline{g}} + \underline{p} \ (\text{mod } n) \mid \underline{p}=[p, \dots, p] %\in \mathbb{Z}^k
%    ,\  p= 0,1, \dots,n-1\},
%\end{equation}
%\end{small}
%the modulo operation ensures that the values wrap around within the range \([0,\ n-1]\).
%We are interested in the generator \(\underline{g} \in \underline{G}\) whose corresponding \(\underline{\mathbf{C}}_{\underline{g}}\) has \(n\) distinct ordered \(k\)-tuples and forms a Hamiltonian cycle, also, the \(\underline{\mathbf{C}}_{\underline{g}}\) must be a subset of \(\underline{\mathbf{A}}_{n,k}\), this implies that whose members must be in valid orientations.
\begin{definition}
%Two valid generators \(\underline{g}\) and \(\underline{g}'\) are distinct if 
Two generators \(\underline{g}\) and \(\underline{g}'\) are said to be distinct iff 
$\underline{\mathbf{C}}_{\underline{g}} \cap\underline{\mathbf{C}}_{\underline{g}'} = \emptyset$ which means that any vector of $\underline{\mathbf{C}}_{\underline{g}}$ is distinct from all the vectors in $\underline{\mathbf{C}}_{\underline{g'}}$.
\end{definition}
%From the adopted ordering of $\underline{\mathbf{A}}_{n,k}$ and from \eqref{eq:Cg}, we know that two distinct generator are invariant under cyclic shifts.
Here we need to clarify that the above $\underline{\mathbf{C}}_{\underline{g}}$ and $\underline{\mathbf{C}}_{\underline{g}'}$ need not be subsets of $\underline{\mathbf{A}}_{n,k}$. In such case, we use $|\underline{\mathbf{C}}_{\underline{g}}|$ to represent the number of distinct entries in $\underline{\mathbf{C}}_{\underline{g}}$. 

We also have the following definition.
\begin{definition}   
The \emph{period} of a generator \(\underline{g}\) is $|\underline{\mathbf{C}}_{\underline{g}}|$, corresponding to the number of distinct elements of $\underline{\mathbf{C}}_{\underline{g}}$. 
\end{definition}
From the above, we see that the period of \(\underline{g}\) is the smallest integer \(p \in [n]\) such that the vector \(\underline{r}_{\underline{g}} + p\cdot \underline{1}\) (mod \(n\)) is a cyclic shift of \(\underline{r}_{\underline{g}}\).

At this point, it is easy to show the following. %The generator \(\underline{g}\) defines the differences between consecutive components in the representative \(\underline{r}_{\underline{g}} = [r_1, r_2, \dots, r_k]\). We assume that \(\underline{r}_{\underline{g}}\) is in increasing order, i.e., \(r_1 = 1 < r_2 < \dots < r_k\). Consequently, the differences \(g_1, g_2, \dots, g_{k-1}\) must satisfy \(g_i \geq 1\) for all \(i\). To understand \(\underline{g}\)'s period, we arrange the digits of the corresponding representative \([r_1, r_2, \dots, r_k]\) on a circle (Figure \ref{fig:cycle}). Each time all the digits are incremented by 1 (mod \(n\)), the configuration of \(r_i\)s rotates clockwise around the circle.
% \begin{itemize}
%     \item If \(g_k\) is negative (\(\sigma \leq \lfloor n/2 \rfloor\)), the components \(r_1, r_2, \dots, r_k\) are clustered within half the circle. As a result, \(n\) full rotations are required to return to the original configuration, making the period \(n\).
% \end{itemize}
\begin{proposition}
\label{pro:period}
%   Period and Coprimality: \\
If \(n\) and \(k\) are coprime, then the generator 
\(\underline{g} = [g_1, g_2, \dots, g_k]\)
%\ g_{i}\geq 1, i\in[k-1]\),
has period $n$.
%   \(\underline{g}\), where \(g_1, g_2, \dots, g_{k-1}\geq 1\) is \(n\).
% and \sigma \leq \sigma_max 
\end{proposition}
\begin{proof}
For generator \(\underline{g} = [g_1, g_2, \dots, g_k]\), recall that the representative $ 
    \underline{r}_{\underline{g}} = [1, 1+g_1, \dots, 1+g_1+\dots+g_{k-1}]$ with $i$th entry $r_i = 1+\sum_{j=1}^{i-1} g_{j}$, yields $
        \underline{\mathbf{C}}_{\underline{g}} = \{\underline{r}_{\underline{g}} + p\cdot \underline{1} \mod n \mid p \in \{0, 1, \dots, n-1\}\}.$
    If \(|\underline{\mathbf{C}}_{\underline{g}}| = d < n\), then \([r_1+d, r_2+d, \dots, r_k+d]\) is a permutation of \(\underline{r}_{\underline{g}}\), and \(d\) must divide \(n\), which means that \(n = k_1 d\) for some integer \(k_1\). 
    Next, consider the arrangement of \(r_1, r_2, \dots, r_k\) on the circle. Divide the circle into \(d\) equal segments of size \(k_1 = \frac{n}{d}\), and note that to maintain periodicity, the pattern of \(r_1, r_2, \dots, r_k\) must appear similarly in each of the \(d\) segments. This implies that \(d\) divides \(k\), which contradicts the fact that \(n\) and \(k\) are coprime, which means that the period \(|\underline{\mathbf{C}}_{\underline{g}}|\) is $n$.    
\end{proof}
%For when \(n\) and \(k\) are prime, and \(k < \frac{n}{2}\), we will further demonstrate that the \(\underline{\mathbf{C}}_{\underline{g}}\) also forms a Hamiltonian cycle. Before this, we need the following definition. 

% we automatically have 1-n,k,k design.

\begin{figure}[htbp]

  \centering
  \includegraphics[width=0.3\textwidth]{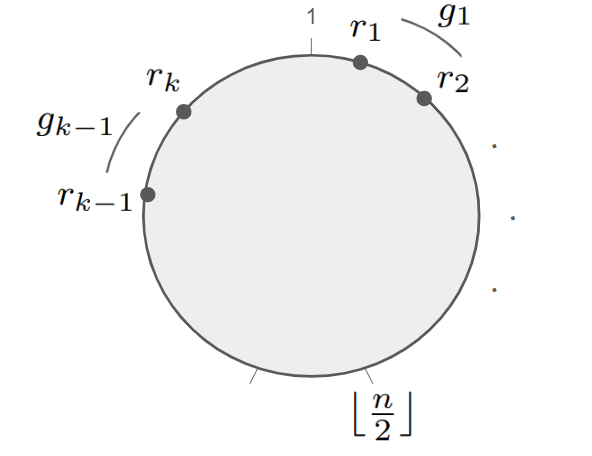}
  \caption{A configuration of the representative \(\underline{r}_{\underline{g}} = [r_1, r_2, \dots, r_k]\). 
  The corresponding generator \(\underline{g}\) indicates the difference between consecutive entries of \(\underline{r}_{\underline{g}}\), meaning that \(g_1=r_2-r_1, \ \dots,\ g_{k-1}=r_k-r_{k-1}\).
  The period indicates the minimum number of clockwise cyclic shifts required to return to the starting configuration.}
  \label{fig:cycle}
\end{figure}

\subsection{Designing Generators}
We aim to create a generator set \(\underline{\mathbf{G}}_{\text{tot}}\) containing all the generators \(\underline{g}_1, \ldots, \underline{g}_N\), such that the corresponding vector sets $\underline{\mathbf{U}}_1 = \underline{\mathbf{C}}_{\underline{g}_1},\dots,\underline{\mathbf{U}}_N = \underline{\mathbf{C}}_{\underline{g}_N}$ partition $\underline{\mathbf{A}}_{n,k}$ (guaranteeing that the different $\underline{\mathbf{C}}_{\underline{g}_i}$ are disjoint and cover $\underline{\mathbf{A}}_{n,k}$) and such that the corresponding cycles $\boldsymbol{\mathcal{S}}_{\underline{g}_1},\dots,\boldsymbol{\mathcal{S}}_{\underline{g}_N}$ (see \eqref{eq:cycleS1}) are Hamiltonian.  
In the following, we will explicitly identify these $N$ distinct generators. To proceed, let us consider parameters $\sigma_{\text{min}}, \sigma_{\text{max}}$ which we respectively set equal to $\sigma_{\text{min}} = k-1,\ \sigma_{\text{max}} = n - \left\lceil \frac{n}{k} \right\rceil$.  Then for all $\sigma \in[\sigma_{\text{min}},\sigma_{\text{max}}]$, we form 
\begin{small}
\begin{equation}
\label{eq:S_sigma}
\underline{\mathbf{S}}_{\sigma} = \{\underline{c}=[c_{1}, \dots, c_{k-1}] 
\mid \sum_{i=1}^{k-1} c_{i} = \sigma, \ 1\leq c_i \leq n-\sigma\}
\end{equation}
\end{small}
and then we form 
\begin{small}
\begin{equation} \label{eq:Gsigma}
\underline{\mathbf{G}}_{\sigma} = 
%\underline{G}_{\sigma, 1} = 
\bigg\{[\underline{c}, m] : \underline{c} \in \underline{\mathbf{S}}_{\sigma}, \ 
m = 
\begin{cases} 
-\sigma, & \text{if } \sigma \leq \lfloor\frac{n}{2}\rfloor, \\
n - \sigma, & \text{otherwise}
\end{cases}\ \bigg\}
\end{equation}
\end{small}
yielding generator set 
\begin{small}
   \begin{equation}
       \underline{\mathbf{G}}_{\text{tot}}=\bigcup_{\sigma = \sigma_{\text{min}}}^{\sigma_{{\text{max}}}} \underline{\mathbf{G}}_{\sigma}.
   \end{equation} 
\end{small}
%Now, all the generator in \(\underline{g} \in \underline{G}_{\text{class 1}}\) are invariant under cyclic shifts, so all of them are distinct. 

\subsection{Proof of Hamiltonian Decomposition}
We now show that $\underline{\mathbf{G}}_{\text{tot}}$ yields a Hamiltonian decomposition. We begin with the following lemma. 
\begin{lemma} \label{lem:GgivesOnlyHamiltonCycles}
For each generator $\underline{g} \in \underline{\mathbf{G}}_{\text{tot}}$, the corresponding \(\boldsymbol{\mathcal{S}}_{\underline{g}}\) is a Hamiltonian cycle.
\end{lemma}
\begin{proof}
We first note that due to~\eqref{eq:S_sigma} and \eqref{eq:Gsigma}, and due to having $n$ prime as well as having the first entry $g_1$ satisfy $g_1<n$, it is the case that \((1,1+g_1, \dots, 1+(n-1)g_1)\) must be a permutation of \((1,2, \dots,n)\), and thus that \(\{\underline{r}_{\underline{g}}, \underline{r}_{\underline{g}}+g_1\cdot \underline{1}, \dots, \underline{r}_{\underline{g}}+(n-1)g_1\cdot \underline{1}\}\) is equal to \(\underline{\mathbf{C}}_{\underline{g}}=\{\underline{r}_{\underline{g}}+p\cdot\underline{1} \mid p \in\{0,1,\ldots,n-1\}\}\). From Proposition \ref{pro:period}, we know that the period of \(\underline{\mathbf{C}}_{\underline{g}}\) is \(n\), and thus that the elements of \( \{\underline{r}_{\underline{g}}, \underline{r}_{\underline{g}}+g_1\cdot \underline{1}, \dots, \underline{r}_{\underline{g}}+(n-1)g_1\cdot \underline{1}\}\) are distinct, which means that \(\boldsymbol{\mathcal{S}}_{\underline{g}}=\{1, \underline{r}_{\underline{g}}, 1+g_1, \underline{r}_{\underline{g}}+g_1.\underline{1}, 1+2g_1, \dots, 1+ng_1, \underline{r}_{\underline{g}}+(n-1)g_1.\underline{1}, 1 \} \) visits all vertices from $1$ to $n$, and is thus a Hamiltonian cycle.
\end{proof}

We now proceed with the main result. 
\begin{theorem} \label{trm:FinalTheorem}
The generators $\underline{g}_1,\dots,\underline{g}_N$, $N =\frac{{n \choose k}}{n}$, in $\underline{\mathbf{G}}_{\text{tot}}$ are distinct, and the partition $\big\{\boldsymbol{\mathcal{S}}_{\underline{g}_i}, \ i\in[N] \big\}$ forms a Hamiltonian decomposition of $\underline{\mathbf{A}}_{n,k}$.\\
\end{theorem}

\begin{proof}
%We will here show that any $k$-tuple in $\underline{\mathbf{A}}_{n,k}$ is assigned to one generator $\underline{g} \in \underline{\mathbf{G}}_{\text{tot}}$. This will show that $\big\{  \underline{\mathbf{C}}_{\underline{g}}, \ \underline{g} \in \underline{\mathbf{G}}_{\text{tot}} \big\}$ covers $\underline{\mathbf{A}}_{n,k}$, and then by definition of distinct generators and of Hamiltonian cycles, we can conclude that $\underline{\mathbf{G}}_{\text{tot}}$ consists of \(\frac{{n \choose k}}{n}\) distinct generators. 
Let us link an arbitrary \(\underline{a}=[a_1,a_2,\dots,a_k] \in \underline{\mathbf{A}}_{n,k}\) to its unique and identified generator $\underline{g}_1,\underline{g}_2,\dots$ from \(\underline{\mathbf{G}}_{\text{tot}}\). Let $\underline{s}_{\underline{a}}=[s_1, s_2, \dots, s_{k-1}]\in \mathbb{Z}_{+}^{k-1}$, $s_i=d_{ro}^n(a_{i+1},a_{i})$, be the corresponding difference vector, where $d_{ro}^n(y,x) = y-x, \ y\geq x$, and where $d_{ro}^n(y,x)  = n-(x-y)$ otherwise. Now let $\gamma \triangleq s_1+ ...+s_{k-1}$, and let us recall that due to the adopted ordering, we have \(\norm{\underline{s}_{\underline{a}}}_{\infty}\leq n-\gamma \). We consider the following cases, and for each case, we will identify the parameters $\sigma,m$ in \eqref{eq:Gsigma} that correspond to $[\underline{s}_{\underline{a}} \ m]$ and thus to $\underline{a}$. In other words, for each $\underline{a}$ we will identify the unique distinct generator, from the final set of $N$ distinct generators, that will have $\underline{a}$ as its hyperedge. 

\emph{Case 1:} $\gamma \leq \lfloor\frac{n}{2}\rfloor$.   Here, after setting $m=-\gamma$, we see from \eqref{eq:Gsigma} that $\underline{g} = [s_1, s_2, ..., s_{k-1},m] \in \underline{\mathbf{G}}_{\sigma}$ for $\sigma=\gamma$. This $ \underline{g} $ belongs in the set of $N$ distinct generators, because any rotation of $\underline{g} $ would entail a negative entry other than the last, which would violate~\eqref{eq:Gsigma}.

Case 2: $\lfloor\frac{n}{2}\rfloor+1 \leq\gamma \leq \lfloor\frac{n}{2}\rfloor+ \lfloor\frac{k}{2}\rfloor -1$. We can see that for any $\underline{a} \in \underline{\mathbf{A}}_{n,k}$, the vector \(\underline{g} = [\underline{s}_{\underline{a}} , m] = [s_1, \dots, s_{k-1}, m]\), where \(m=n-\gamma\), is in the set generated in \eqref{eq:Gsigma} when \(\sigma=\gamma\).  This $ \underline{g} $ again belongs in the set of $N$ distinct generators, because any rotation of $\underline{g} $ would violate the condition in~\eqref{eq:Gsigma} which states that the largest entry must appear at the end of the vector. To see that indeed all rotations of the above $\underline{g} $ would violate this condition, we see the following. First we note that for our $\underline{s}_{\underline{a}} = [s_1, \dots, s_{k-1}]$, it must be the case that \(s_i < m, \ i=1,\dots,k-1\) because if there existed an index $i\in[k-1]$ such that \(s_i=m\), it would imply that \(s_1+\ldots+s_{k-1}=\gamma\geq m+(k-2)=n-\gamma+k-2\), because $s_i\geq m$ for some $i$, and for the rest we have $s_i\geq 1$. Thus we would have $\gamma\geq m+(k-2)=n-\gamma+k-2$ which implies that \(\gamma\geq \frac{n+k-2}{2}=\lfloor\frac{n}{2}\rfloor+ \lfloor\frac{k}{2}\rfloor\), because $k,n$ are odd. This inequality though contradicts the defining region of $\gamma$ for this case. Thus since \(s_i < m, \ i=1,\dots,k-1\), we can conclude that any rotation of $[s_1, s_2, ..., s_{k-1},m]$ would violate the constructive condition in~\eqref{eq:S_sigma} and~\eqref{eq:Gsigma}. Thus the described generator $\underline{g}$ -- for each of the $\underline{a} $ for this case -- belongs in the set of $N$ distinct generators.

Case 3: $ \lfloor\frac{n}{2}\rfloor+ \lfloor\frac{k}{2}\rfloor \leq \gamma \leq n -\lceil \frac{n}{k} \rceil$. For this case, the design in~\eqref{eq:S_sigma} and~\eqref{eq:Gsigma} corresponds to \(\sigma=\gamma\) and to \(m=n-\gamma\).  First note that having $n-\gamma< \norm{\underline{s}_{\underline{a}}}_\infty$ contradicts the ordering condition on \(\underline{a}\), and thus we conclude that $n-\gamma\geq \norm{\underline{s}_{\underline{a}}}_\infty$. Consider the following two sub-cases. Case 3a): When $n-\gamma >\norm{\underline{s}_{\underline{a}}}_\infty $, the vector $[s_1, s_2, ..., s_{k-1}, m=n-\gamma]$ is a chosen generator because any rotation would violate the condition that the largest entry be at the end, and thus would violate the conditions in~\eqref{eq:S_sigma} and~\eqref{eq:Gsigma}.  Case 3b): Now suppose $n-\gamma =\norm{\underline{s}_{\underline{a}}}_\infty $, and consider vector \(\underline{g}=[s_1, s_2, ...,s_{k-1}, n-\gamma]\), and let $k'$ denote the number of entries of \(\underline{g}\) that are equal to \(\norm{\underline{s}_{\underline{a}}}_\infty\). Naturally \(k'\geq 2\) (corresponding to the last entry and at least one more). Since \(k'\geq 2\), there is at least a cyclic version of \(\underline{g}\) which creates the same \(\underline{\mathbf{C}}_{\underline{g}}\). Considering all \(k\) cyclic shifts of \(\underline{g}\), we see that since \(\underline{g}\) has \(k'\) entries equal to \(\norm{\underline{s}_{\underline{a}}}_\infty\), then only \(k'\) of these shifts (including the identity) can be selected because one of these $k'$ maximum-valued entries must be the last entry. We have $k'$ candidates, and one of the will be selected. To canonically select, we create the ordered sets $\underline{h}_i,i\in[k']$, and we select the shift that has the highest lexicographic ordering. Only one candidate is chosen, canonically, in the set of $N$ distinct generators.
%At the same time though, either \(\underline{g}\) or one of its $k'-1$ cyclic shifts, must be associated to the original $\underline{a}$. Thus, the position of the digits with maximum value matters. Suppose \(k'\)-sized ordered sets \(\underline{h}_1,\underline{h}_2, \ldots, \underline{h}_{k'}\) determine the location of \(k'\) cyclic version of \underline{g}. For example, for \(k=5\), \(n=17\) and \(\underline{g}=[4,3,2,4,4]\), we \(k'=3\), we have \(k'=3\) shifted version \(\underline{g}_{\text{shifted1}}=[4,4,3,2,4], \underline{g}_{\text{shifted2}}=[3,2,4,4,4],\underline{g}_{\text{shifted3}}=[4,3,2,4,4] \) and respectively \(\underline{h}_1=[1,2,5],\underline{h}_2=[3,4,5],\underline{h}_{k'=3}=[1,4,5]\). We sort \(\underline{h}_1,\underline{h}_2, \ldots, \underline{h}_{k'}\) based on lexicographical order and choose the biggest one. Its corresponding \(\underline{g}_{\text{shifted}}\) will be the generator.

Case 4: $\gamma > n -\lceil \frac{n}{k} \rceil $. 
Having $\gamma > n -\lceil \frac{n}{k} \rceil $ implies that $s_i >  n-\gamma$ for some $i\in[k-1]$. If not, i.e., if $s_i\leq  n-\gamma, \forall i\in[k-1]$, then we would have that  \(s_1+\ldots+s_{k-1}=\gamma \leq (k-1)(n-\gamma)\) which would imply that \(k \gamma \leq (k-1)n\) and that \(\gamma \leq \lfloor \frac{(k-1)n}{k}\rfloor=n -\lceil\frac{n}{k}\rceil\) which violates the $\gamma$ region of the case. Thus we are left with the case of $s_i >  n-\gamma$ for some $i\in[k-1]$, which contradicts with the ordering adopted. Thus no $\underline{a} \in \underline{\mathbf{A}}_{n,k}$ falls under this case. This justifies setting $\sigma_{\text{max}} = n -\lceil \frac{n}{k} \rceil$ in~\eqref{eq:S_sigma} and~\eqref{eq:Gsigma}.

From the above, we now know that every ordered \(k\)-tuple \(\underline{a}=[a_1,\ a_2,\ \dots,\ a_k]   \in \underline{\mathbf{A}}_{n,k} \), is associated to a unique generator \(\underline{g}\in \underline{\mathbf{G}}_{\text{tot}}\), and thus we know that $\underline{\mathbf{A}}_{n,k}$ is covered. We also know from Lemma~\ref{lem:GgivesOnlyHamiltonCycles} that each such $\underline{g}$ yields a Hamiltonian cycle $\boldsymbol{\mathcal{S}}_{\underline{g}}$ which has the aforementioned $\underline{a}$ as a hyperedge. Now let us consider a bipartite graph where on the left we have the vertices $\mathcal{L}$ representing the distinct generators $\underline{g}_1,\underline{g}_2,\dots,\underline{g}_{|\mathcal{L}|}$ from \(\underline{\mathbf{G}}_{\text{tot}}\), and on the right we have the vertices $\mathcal{R}$ corresponding to the \(k\)-tuples in \(\underline{\mathbf{A}}_{n,k}\). Since every \(\underline{a}  \in \underline{\mathbf{A}}_{n,k} \) corresponds to a single generator, the degree of a node in $\mathcal{R}$ is fixed at $\deg(\mathcal{R})$ = 1.   On the other hand, directly from Proposition~\ref{pro:period}, we know that each generator from \(\underline{\mathbf{G}}_{\text{tot}}\) has period \(n\), and is thus connected to \(n\) different \(k\)-tuples in $\underline{\mathbf{A}}_{n,k}$ ($n$ nodes in $\mathcal{R}$), which implies that the degree of a node in $\mathcal{L}$ is fixed at $\deg(\mathcal{L}) = n$. Let us now recall that $|\mathcal{R}|=|\underline{\mathbf{A}}_{n,k}|={n \choose k}$, and let us recall from basic combinatorics that $|\mathcal{L}| \cdot \deg(\mathcal{L})=|\mathcal{R}|\cdot \deg(\mathcal{R})$, which means that $|\mathcal{L}|=\frac{{n \choose k}}{n} = N$. These $N$ generators yield $N$ distinct Hamiltonian cycles $\{\boldsymbol{\mathcal{S}}_{\underline{g}_1},\dots,\boldsymbol{\mathcal{S}}_{\underline{g}_N}\}$. The proof is complete. 
\end{proof}

The following remark may be of use. 
\begin{remark}
We here clarify that $\underline{\mathbf{G}}_{\text{tot}}$, as generated in~\eqref{eq:Gsigma}, may contain more than $N$ entries. From these, only $N$ will be distinct, and we are able to identify these $N$ distinct generators \emph{explicitly}. Equivalently, for each $\underline{a} \in \underline{\mathbf{A}}_{n,k}$, we identify explicitly the one unique generator $\underline{g} \in \underline{\mathbf{G}}_{\text{tot}}$ (and thus the unique cycle $\boldsymbol{\mathcal{S}}_{\underline{g}}$) that has that specific $\underline{a}$ as a hyper-edge. The ability to explicitly identify and construct the set of $N$ distinct generators depends on the mathematical expressions in~\eqref{eq:S_sigma} and~\eqref{eq:Gsigma}, the condition that defines the adopted ordering, and a basic lexicographic ordering. 
\end{remark}
We proceed with an example. 

\subsection{Example of a New Construction}
We here give an example for a newly constructed case, for $n=11,k=5$. For brevity, we will here describe the first vector (the representative $\underline{r}_{\underline{g}}$) of each Hamiltonian cycle $\boldsymbol{\mathcal{S}}_{\underline{g}}$. We provide $N = {11 \choose 5}/{11} = 42$ distinct generators. As \(\sigma_{\text{min}}=k-1=4\) and \(\sigma_{\text{max}}=8\), we draw our generators from the following sets below. %We need to provide $N = \frac{{11 \choose 5}}{11} = 42$ distinct generators from \(\underline{\mathbf{G}}_{\text{tot}}=\bigcup_{\sigma=4}^{8}\underline{\mathbf{G}}_{\sigma}\), which we list below.
\begin{small}
\begin{align*}
\underline{\mathbf{G}}_4 &= \bigg\{[1,1,1,1,-4]\bigg\},\\
\underline{\mathbf{G}}_5 & = \bigg\{[2,1,1,1,-5],[1,2,1,1,-5],[1,1,2,1,-5],[1,1,1,2,-5]\bigg\},\\
\underline{\mathbf{G}}_6 &= \bigg\{[3,1,1,1,5],[1,3,1,1,5],[1,1,3,1,5],[1,1,1,3,5], \\
                    & [2,2,1,1,5],[2,1,2,1,5],[2,1,1,2,5],[1,2,2,1,5],[1,2,1,2,5],\\
                    & [1,1,2,2,5]\bigg\},\\
\underline{\mathbf{G}}_7 &=\bigg\{[1,1,1,4,4],[1,1,4,1,4],[3,2,1,1,4],[3,1,2,1,4],\\
      & [3,1,1,2,4],[1,3,2,1,4],[1,3,1,2,4],[2,3,1,1,4],[1,1,3,2,4],\\
      & [1,2,3,1,4],[2,1,3,1,4],[1,1,2,3,4],[1,2,1,3,4],[2,1,1,3,4],\\
      & [1,2,2,2,4],[2,1,2,2,4],[2,2,1,2,4],[2,2,2,1,4]\bigg\},\\
 \underline{\mathbf{G}}_8 &= \bigg\{[1,2,2,3,3],[2,1,2,3,3],[2,2,1,3,3],[1,2,3,2,3],\\
 & [2,1,3,2,3],[2,2,3,1,3],[1,1,3,3,3],[1,3,1,3,3],[2,2,2,2,3]\bigg\}.
 \end{align*}
\end{small}
For \(\underline{g}_1=[g_1, g_2, g_3, g_4, g_5]=[1,1,1,1,-4]\), the representative vector is
\[\underline{r}_{\underline{g}_1}=[1+g_1, \dots, 1+g_1+g_2+g_3+g_4]=[1,2,3,4,5]\] giving us 
\begin{small}
\begin{align*}
\underline{\mathbf{C}}_{\underline{g}_1}&=\bigg\{[1,2,3,4,5], [2,3,4,5,6], \dots, [6,7,8,9,10],[7,8,9,10,11], \\
&[8,9,10,11,1] , [9,10,11,1,2], [10,11,1,2,3], [11,1,2,3,4]\bigg\}
\end{align*}
\end{small}
which yields the Hamiltonian cycle 
\begin{small}
\begin{align*}
\boldsymbol{\mathcal{S}}_{\underline{g}_1} & =\bigg\{1, [1,2,3,4,5], 2, [2,3,4,5,6], 3 ,[3,4,5,6,7], 4,[4,5,6,7,8],\\
& 5, [5,6,7,8,9], 6,[6,7,8,9,10], 7, [7,8,9,10,11],8, [8,9,10,11,1] \\
& ,9, [9,10,11,1,2], 10, [10,11,1,2,3], 11, [11,1,2,3,4], 1\bigg\}.
\end{align*}
\end{small}
The other Hamiltonian cycles are computed in the same manner, using the remaining $41$ generators. 

\section{Conclusions}  
This work advances the field of Hamiltonian decompositions by presenting explicit designs for complete \(k\)-uniform hypergraphs \(K_n^k\) for all \(k\) and \(n\) prime, addressing a gap left by previous existence-based methods and restricted explicit constructions. The broad applicability of these decompositions in distributed computing, coded caching, resource allocation, load balancing, and fault tolerance underscores their fundamental role in optimizing modern networks and systems, as well as underscores the utility of explicit designs.  Future directions may include extending these constructions to composite \(n\) and exploring further connections with emerging technologies in caching and network optimization. 
\section*{Acknowledgment}
This work is supported by the European Research Council (ERC) through the ERC-PoC Grant 101101031, and the Eurecom Huawei Chair.

%% To balance the columns at the last page of the paper use this
%% command:
%%
%\enlargethispage{-1.2cm} 
%%

\bibliographystyle{IEEEtran}
\bibliography{references}
\enlargethispage{-1.2cm} 
\end{document}